\documentclass{amsart} 
\begin{document} 
\sloppy
\title{On q-deformed quantum stochastic calculus}
\author{Piotr \'Sniady}
\address{Institute of Mathematics, \\ University of Wroclaw, \\
pl.~Grunwaldzki 2/4, 50-384 Wroc\l{}aw, Poland} 
\email{psnia@math.uni.wroc.pl}
\begin{abstract}
In this paper we investigate a quantum stochastic calculus build of  
creation, annihilation and number of particles operators which fulfill some 
deformed commutation relations. 

Namely, we introduce a deformation of a 
number of particles operator which has simple commutation relations with 
well known $q$-deformed creation and annihilation operators. Since all 
operators considered in this theory are bounded we do not deal with some 
difficulties of a non deformed theory of Hudson and Parthasarathy \cite{HuP}.
We define stochastic integrals and estimate them in the operator norm.
We prove It\^o's formula as well. 
\end{abstract}
\maketitle
\newcommand{\cal}[0]{\mathcal} 
\newcommand{\jed}[0]{{\bf 1}}
\newcommand{\Id}[0]{{\rm Id}}
\newcommand{\ksi}[0]{\xi}
\newcommand{\fin}[0]{_{\rm fin}}
\newcommand{\zero}[0]{{\bf 0}}
\newcommand{\Pn}[0]{P^{(n)}}
\newcommand{\free}[0]{_{\rm free} }
\newcommand{\N}[0]{{\mathbb N}}
\newcommand{\Z}[0]{{\mathbb Z}}
\newcommand{\C}[0]{{\mathbb C}}
\newcommand{\Hn}[0]{{\cal H}^{\otimes n}}
\newcommand{\Ha}[0]{{\cal H}}
\newcommand{\Ka}[0]{{\cal K}}
\newcommand{\El}[0]{{\cal L}}
\newcommand{\lamjed}[0]{\lambda}
\newcommand{\Rplus}[0]{{\mathbb R}_+}
\newcommand{\Hado}[1]{\Ha^{\otimes #1}}
\newcommand{\gwia}[0]{^{\ast}}
\newcommand{\dagg}[0]{^{\dagger}}
\newcommand{\A}[0]{{\cal A}}
\newcommand{\op}[0]{} 
\newcommand{\zeruj}[0]{\setcounter{subsection}{1} \setcounter{lem}{0}}
\newcommand{\lamb}[0]{^{\lambda}}
\newtheorem{lem}{Lemma}[section]
\newtheorem{theo}[lem]{Theorem}
\newtheorem{defin}[lem]{Definition}
\newtheorem{prop}[lem]{Proposition}
\newtheorem{coro}[lem]{Corollary}
\section{Introduction}
The aim of this paper is to construct a quantum stochastic calculus in which 
all operators are bounded and which would unify classical examples we 
mention below.
\subsection{Classical examples of quantum stochastic calculi}
The fundamental observation which inspired the development of 
Hudson--Parthasarathy stochastic calculus \cite{HuP} was that a family of 
commuting selfadjoint operators and a state $\tau$ induce (by the 
spectral theorem) measures which can be interpreted as joint distributions 
of a certain stochastic process.

The most important examples are $B(t)=A(t)+A\gwia(t)$ which corresponds to 
the Brownian motion and $P_l (t)=\sqrt{l}\ B(t)+\Lambda(t)+lt \jed$ which 
correspond to the Poisson process with intensively $l$.  
$A(t),A\gwia(t),\Lambda(t)$ ($t\geq 0$) called annihilation, creation and 
gauge process have values being unbounded operators acting on some 
Hilbert space called bosonic Fock space. 

For all $s,t\geq 0$ they fulfill the following commutation relations:
\begin{equation}\label{eq:komut1}[A(t),A(s)]=[A\gwia(t),A\gwia(s)]=0, 
\end{equation} \begin{equation}\label{eq:komut1a} 
[\Lambda(t),\Lambda(s)]=0,\end{equation} 
\begin{equation}\label{eq:komut2}[A(t),A\gwia(s)]=\min(t,s) 
\jed,\end{equation} \begin{equation}\label{eq:komut3}[A(t), 
\Lambda(s)]=A[\min(t,s)],\end{equation}
\begin{equation}\label{eq:komut4}[\Lambda(t),A\gwia(s)]=A\gwia[\min(t,s)]. 
\end{equation}
In the Fock space there exists a unital cyclic  vector $\Omega$ such 
that $A(t)\Omega=0$. The state $\tau$ is defined as 
follows: $$\tau(S)=\langle\Omega,S\Omega\rangle.$$

Stochastical integrals with respect to the Brownian motion or Poisson 
process can therefore be written as integrals with respect to creation, 
annihilation and gauge processes. A stochastic calculus in which such 
integrals are considered was constructed by Hudson and Parthasarathy 
\cite{HuP}. However the fact that operators considered in this theory are 
unbounded causes serious technical problems. For example equations 
(\ref{eq:komut1})--(\ref{eq:komut4}) can be treated only informally and has 
to be clarified in a more complicated way. Moreover, a product of two 
stochastic integrals (considered in It\^o's formula) is not well--defined 
and has to be evaluated in the weak sense.

A second important example is a fermionic stochastic calculus 
\cite{BSW1,BSW2} in which in equations (\ref{eq:komut1})--(\ref{eq:komut4}) 
commutators were replaced by anticommutators.

The third group of examples is connected with free probability 
in which the notion of classical independence of random 
variables was replaced by a non commutative notion of freeness. 
Biane and Speicher \cite{BiS} considered integrals with respect to the free 
Brownian motion what is a generalization of the It\^o's integral. 
On the other hand the approach of K\"ummerer and Speicher \cite{KuS} is 
rather relate to the calculus of Hudson and Parthasarathy: a free 
Brownian motion is represented as a family of non commuting selfadjoint 
operators $B(t)=A(t)+A\gwia(t)$ ($t\geq 0$) where $A(t), A\gwia(t)$ fulfill 
only a relation \begin{equation}\label{eq:free} A(t) 
A\gwia(s)=\min(t,s)\jed,\end{equation} 
for all $t,s\geq 0$ and a state $\tau$ is defined as 
$\tau(S)=\langle\Omega,S\Omega\rangle$ for a unital cyclic vector $\Omega$ 
such that $A(t)\Omega=0$. Stochastic integrals are evaluated with respect to 
$A(t)$ and $A\gwia(t)$ separately.

\subsection{Overview of this paper}
In order to avoid problems of Hudson and Parthasarathy's theory we 
postulate that all operators considered in our stochastic calculus should be 
bounded. Therefore we shall replace commutation relations of 
Hudson--Parthasarathy's calculus by some deformed analogues. 

We start with the $q$-deformed commutation relation which was postulated by 
Frisch and Bourret \cite{FB}:
\begin{equation} \label{eq:qdeformed} a(\phi) a\gwia(\psi)=q a\gwia(\psi) 
a(\phi)+\langle \phi,\psi \rangle, \end{equation}
for all $\phi,\psi\in\Ha$, where $\Ha$ is a Hilbert space, $a(\phi)$ called 
annihilation operator and its adjoint $a\gwia(\phi)$ called creation 
operator are operators acting on some Hilbert space $\Gamma_\Ha$.

If in the equation (\ref{eq:qdeformed}) we take $q=1$ we obtain bosonic 
commutation relation (\ref{eq:komut2}), for $q=-1$ we obtain fermionic 
anticommutation relation and for $q=0$ we obtain free relation 
(\ref{eq:free}), therefore $q$-deformed commutation relation unifies these 
three basic cases. 

In the section \ref{sec:deformed} we shall repeat Bo\.zejko and 
Speicher's \cite{BoS} construction of $q$-deformed Fock space 
$\Gamma_\Ha$ and bounded operators $a(\phi)$, $a\gwia(\phi)$ which fulfill 
(\ref{eq:qdeformed}). Furthermore we construct a bounded operator 
$\lambda_\mu$ which acts on $\Gamma_\Ha$ and is a deformation of 
Hudson--Parthasarathy gauge operator and an auxiliary operator 
$\gamma_\mu:\Gamma_\Ha\rightarrow\Gamma_\Ha$ which is a deformation of the 
identity.
We show commutation relations fulfilled by these operators. It turns out 
that these commutation relations allows us to write any product of these 
operators in a special order which is a generalization of Wick or normal 
ordering.

In the section \ref{sec:calki} we define stochastic integrals with respect 
to four basic processes: annihilation $A(t)$, creation $A\gwia(t)$, gauge 
$\Lambda_\mu(t)$ and time T(t). 
Since in the non commutative probability the integrand does not commute 
with the increments of integrator, we have to decide if the 
integrand should be multiplied from the left or from the right by the 
integrator. In fact, we shall investigate even a more general case, namely 
after Speicher and Biane \cite{BiS} we consider so called bioperators and 
biprocesses, so that the increments of integrator are multiplied both from 
left and right by the integrand.

Just like in the classical theory we first 
define stochastic integrals of simple adapted biprocesses and 
then by some limit procedure we extend stochastic integrals to 
a more general class of biprocesses.

In the section \ref{sec:iterowane} we show that (under certain
assumptions) an integral of a stochastic process is again an 
integrable stochastic process and that such an iterated 
integral is continuous. 

The section \ref{sec:ito} is devoted to the central point of 
this paper, the It\^o's formula, which can be viewed as an integration by 
parts.

\section{Deformed creation, annihilation and number of particles operators}
\label{sec:deformed}
\subsection{Fock space} 
\label{subsec:fockspace}
Let $\Ha$ be a Hilbert space with a scalar product 
$\langle\cdot,\cdot\rangle$. Elements of $\Ha$ will be denoted by small 
Greek letters: $\phi$, $\psi$,\dots 

We shall denote the standard scalar product on $\Hado{n}$ by $\langle 
\cdot,\cdot \rangle\free$ and call it a free scalar product. $\Ha^{\otimes 
n}$ furnished with this scalar product will be denoted by $\Ha^{\otimes 
n}\free$.
By $\Gamma\free(\Ha)$ or simply $\Gamma\free$ we shall denote 
the direct sum of $\Hn\free$, $n\in\N=\{0,1,2,\dots\}$. The space $\Hado{0}$ 
which appears in this sum is understood as a one dimensional space $\C 
\Omega$ for some unital vector $\Omega$.

If $E:{\cal D}(E)\rightarrow\Gamma\free$ for ${\cal D}(E)\subseteq
\Gamma\free$
is a (possibly unbounded) strictly positive operator, we can introduce a
new scalar product 
$\langle\cdot,\cdot\rangle_E=\langle\cdot,E\cdot\rangle\free$
and a Hilbert space $\Gamma_E(\Ha)$ or simply $\Gamma_E$ which is a
completion of ${\cal D}(E)$ with respect to $\langle\cdot,\cdot\rangle_E$. 
The norm in $\Gamma_E$ will be denoted by $\|\cdot\|_E$.

We choose now a parameter of deformation $q\in(-1,1)$ which will be fixed in 
this paper. 

For $n\in\N$ we introduce after Bo\.zejko and Speicher \cite{BoS} a 
$q$-deformed symmetrization operator $P^{(n)}:\Hado{n} \rightarrow 
\Hado{n}$, which is a generalization of a symmetrization (for $q=1$) and 
antisymmetrization (for $q=-1$) operators: $$\Pn(\psi_1\otimes\cdots\otimes 
\psi_n)=\sum_{\sigma\in S_n} q^{{\rm inv}(\sigma)} 
\psi_{\sigma(1)}\otimes\cdots\otimes \psi_{\sigma(n)}, $$ where ${\rm 
inv}(\sigma)=\#\{(i,j):i,j\in \{1,\dots,n\}, i<j, \sigma(i)>\sigma(j)\}$ is 
the number of inversions in permutation $\sigma$. 

\begin{theo} $\Pn$ is a strictly positive operator.
\end{theo}
\begin{proof} Proof can be found in \cite{BoS}. \end{proof}

By $P:{\cal D}(P)\rightarrow\Gamma\free$ (${\cal 
D}(P)\subset\Gamma\free$) we shall denote a closure of the 
direct sum of $\Pn$ and by $\Gamma$ we shall denote $\Gamma_P$.
Since it does not lead to confusions by
$\langle\cdot,\cdot\rangle$ we shall denote both the scalar product in 
$\Ha$ and the $q$-deformed scalar product 
$\langle\cdot,\cdot\rangle_P$ in the Fock space $\Gamma$ and by
$\|\cdot\|$ both the norm in $\Gamma$ and in $\Ha$. Elements of Fock space 
will be denoted by capital Greek letters: $\Phi$, $\Psi$,\dots 

Since now $\Ha^{\otimes n}$ will denote the tensor power of $\Ha$ furnished 
with $q$-deformed scalar product $\langle\cdot,\cdot\rangle$. 

Let $\Pi_j:\Gamma\rightarrow\Hado{j}$ denote the orthogonal projection on 
$\Hado{j}$.

The state $\tau$ which plays the role of a noncommutative expectation value 
is defined as $\tau(X)=\langle\Omega,X\Omega\rangle$ for 
$X:\Gamma\rightarrow\Gamma$.

\subsection{Operators of creation and annihilation}
For $\phi\in\Ha$ we define action of operators $a(\phi), 
a\gwia(\phi):\Gamma_\Ha\rightarrow\Gamma_\Ha$ on simple tensors as follows:
\begin{equation}a\gwia(\phi) (\psi_1\otimes\cdots\otimes \psi_n)=\phi\otimes
\psi_1\otimes\cdots\otimes \psi_n, \label{eq:definicjaagwia} \end{equation}
\begin{equation}a(\phi)(\psi_1\otimes\cdots\otimes \psi_n)=\sum_{i=1}^n 
q^{i-1} \langle \phi,\psi_i\rangle\ \psi_1\otimes\cdots\otimes 
\psi_{i-1}\otimes \psi_{i+1}\otimes\cdots\otimes 
\psi_n,\label{eq:definicjaa} \end{equation}

\subsection{Number of particles operators}
\label{subsec:number}
Now we need to introduce a deformed analogue of number of 
particles operator known also as gauge operator or differential second 
quantization operator.
We require for this deformation to be a bounded operator and to have 
simple commutation relations with $a(\phi)$ and $a\gwia(\phi)$.

For a bounded operator 
$T:\Ha\rightarrow\Ha$ we are looking for 
$\lambda(T):\Gamma_\Ha\rightarrow\Gamma_\Ha$ which action on simple 
tensors is defined as: 
$$\lambda(T) (\psi_1\otimes\cdots\otimes  
\psi_n)= \sum_{i=1}^n f(n)\ \psi_1\otimes\cdots\otimes \psi_{i-1} \otimes 
T(\psi_i) \otimes \psi_{i+1} \otimes\cdots\otimes \psi_n.$$
Except the factor $f(n)$ this definition coincides with a non deformed 
gauge operator. This factor was added in order to make $\lambda(T)$ a 
bounded operator. As it will be proven in the lemma \ref{lem:rownoscnorm} 
it holds if and only if $\sup_{n\in \N} |f(n)| n<\infty$. 

The choice of $f(n)=\mu^n$ for a complex number $\mu$ ($|\mu|<1$) seems to 
be the easiest solution. Therefore we define 
\begin{equation}\label{eq:wzornalambda} \lambda_\mu(T) 
(\psi_1\otimes\cdots\otimes \psi_n)= \sum_{i=1}^n \mu^n\ 
\psi_1\otimes\cdots\otimes \psi_{i-1} \otimes T(\psi_i) \otimes \psi_{i+1} 
\otimes\cdots\otimes \psi_n.\end{equation}

As we shall see in the section \ref{subsec:komutacja} in order to 
interchange the deformed number of particles operator with creation or 
annihilation operators we need to introduce for $|\mu|\leq 1$ an operator 
$\gamma_\mu:\Gamma_\Ha\rightarrow\Gamma_\Ha$ as follows: 
\begin{equation}\label{eq:definicjagamma} \gamma_\mu 
(\psi_1\otimes\cdots\otimes \psi_n)=\mu^n \psi_1\otimes\cdots\otimes 
\psi_n.\end{equation} This operator is a deformed identity operator and for 
$\mu=1$ is equal to identity.

\begin{theo} \label{theo:skonczone}
For $\phi\in\Ha$, $|\mu|<1$ and a bounded $T:\Ha\rightarrow\Ha$ operators 
$a(\phi),a\gwia(\phi),\lambda_\mu(T)$ and $\gamma_\mu$ are bounded and
$$\|\gamma_\mu\|=1,$$
$$\|a(\phi)\|=\|a\gwia(\phi)\|\leq\frac{\|\phi\|}{\sqrt{1-|q|}},$$
$$\|\lambda_\mu(T)\|\leq \|T\|\ \sup_{n\in\N} n\ |\mu|^n.$$
Operators $a\gwia(\phi)$ and $a(\phi)$ as defined in equations
(\ref{eq:definicjaagwia}) and (\ref{eq:definicjaa}) are  adjoint as the 
notation suggests. Furthermore we have
$$\left[\lambda_\mu(T)\right]\gwia=\lambda_{\bar\mu}(T\gwia)$$ 
$$\gamma_\mu\gwia=\gamma_{\bar\mu}$$ \end{theo}
\begin{proof}
It is obvious that for $|\mu|\leq1$ the operator $\gamma_\mu$ is a 
contraction.

The second inequality will be proven in a more general context in the section
\ref{subsec:nierownosci}.

Since $\Hado{n}$ are mutually orthogonal invariant spaces of 
$\lambda_\mu(T)$ from the lemma \ref{lem:rownoscnorm} follows that 
$$\|\lambda_\mu(T)\|_{\Gamma\rightarrow\Gamma}=\sup_{n\in{\N}} 
\|\lambda_\mu(T)\|_{\Hado{n}\rightarrow\Hado{n}}=$$ $$=\sup_{n\in{\N}} 
\|\lambda_\mu(T)\|_{\Hado{n}\free\rightarrow\Hado{n}\free}\leq  \|T\|\
\sup_{n\in{\N}} n |\mu|^n. $$ 

Proof of the fact that $a(\phi)$ and $a\gwia(\phi)$ are adjoint can be 
found in \cite{BoS}. Proofs of the other two equations are straightforward. 
\end{proof}

\begin{lem} \label{lem:rownoscnorm} Suppose that ${\cal V}_i$ ($i=1,2$) 
are vector spaces. ${\cal V}_i$ furnished with a scalar product 
$\langle\cdot,\cdot\rangle_i$ is a Hilbert space denoted by ${\cal K}_i$.

If $P_i:{\cal K}_i\rightarrow{\cal K}_i$ are strictly positive bounded 
operators we can furnish ${\cal V}_i$ with another scalar product $\langle
\cdot,P_i\cdot\rangle_i$ and the resulting Hilbert spaces we shall denote by 
${\cal K}_i'$.

Then operator norms of $S:{\cal K}_1\rightarrow{\cal K}_2$ and $S:{\cal
K}_1'\rightarrow{\cal K}_2'$ are equal for every operator $S:{\cal
V}_1\rightarrow{\cal V}_2$ such that $SP_1=P_2S$.
\end{lem}
\begin{proof} For any polynomial $f(x)$ we have $Sf(P_1)=f(P_2)S$ therefore 
approximating the square root by polynomials we obtain 
$S\sqrt{P_1}=\sqrt{P_2}\ S$. Note that for $v\in {\cal V}_1$ we have
$$\left\|Sv\right\|_{{\cal K}_2'}=
\left\|\sqrt{P_2}\ Sv\right\|_{{\cal K}_2}=\left\|S \sqrt{P_1}\ 
v\right\|_{{\cal K}_2}\leq$$
$$\leq \|S\|_{{\cal K}_1\rightarrow{\cal K}_2} \ \left\|\sqrt{P_1}\ 
v\right\|_{{\cal K}_1}=
\|S\|_{{\cal K}_1\rightarrow{\cal K}_2}\ \|v\|_{{\cal K}_1'},$$
therefore $$\|S\|_{{\cal K}_1'\rightarrow{\cal K}_2'}\leq \|S\|_{{\cal
K}_1\rightarrow{\cal K}_2}.$$
If in the preceding calculations we replace ${\cal K}_i$ by ${\cal K}_i'$ 
(and vice versa) and replace $P_i$ by $P_i^{-1}$  we obtain the opposite 
inequality. \end{proof}

\subsection{Commutation relations} \label{subsec:komutacja}
\begin{theo} For $\phi,\psi\in\Ha$, a bounded operator $T:\Ha\rightarrow\Ha$ 
and $|\mu|,|\nu|<1$ hold 
\begin{equation} a(\phi) a\gwia(\psi)=q 
a\gwia(\psi) a(\phi)+\langle \phi,\psi \rangle,\label{eq:komutacja1} 
\end{equation} 
\begin{equation} a(\phi) \gamma_\mu=\mu \gamma_\mu 
a(\phi),\end{equation}
\begin{equation} \gamma_\mu a\gwia(\phi)=\mu 
a\gwia(\phi) \gamma_\mu ,\end{equation}
\begin{equation} a(\phi) \lambda_\mu(T)=\mu \lambda_\mu(T) 
a(\phi)+\mu \gamma_\mu a(T\gwia \phi), \label{eq:komutacja4} \end{equation}
\begin{equation} \lambda_\mu(T) a\gwia(\phi)=\mu a\gwia(\phi) 
\lambda_\mu(T)+\mu
a\gwia(T\phi) \gamma_\mu, \label{eq:komutacja5} \end{equation}
\begin{equation} 
\lambda_\mu(T)\gamma_\nu=\gamma_\mu\lambda_\nu(T)= 
\lambda_{\mu \nu}(T).\label{eq:komutacja6} \end{equation}
\begin{equation} \gamma_\mu \gamma_\nu=\gamma_{\mu \nu} 
\label{eq:komutacja7} \end{equation}
If bounded operators $T_1,T_2:\Ha\rightarrow\Ha$ commute then
\begin{equation} \lambda_\mu(T_1) \lambda_\nu(T_2)=\lambda_\nu(T_2) 
\lambda_\mu(T_1).\label{eq:komutacja8} \end{equation}
\end{theo}
\begin{proof} Proof is straightforward, we shall omit it. \end{proof}
Since $\gamma_1$ is equal to identity we see that 
in the limit $q,\mu\rightarrow 1$ relations (\ref{eq:komutacja1}), 
(\ref{eq:komutacja4}), (\ref{eq:komutacja5}) and 
(\ref{eq:komutacja6}) correspond to non deformed relations 
(\ref{eq:komut2}), (\ref{eq:komut3}), (\ref{eq:komut4}) and 
(\ref{eq:komut1a}).
Note that contrary to the non deformed case among these commutation 
relations there is none which would allow us to interchange the order of
adjacent two creation or two annihilation operators.

\subsection{Algebra $\A$.}
Suppose $\Ha=\Ka\oplus\Ka^{\perp}$ and $\Ka^{\perp}$ is an 
infinite--dimensional, separable Hilbert space.
We denote by ${\A}_{{\rm fin}\Ka}(\Ha)$ an algebra of bounded 
operators acting on $\Ha$ generated by operators $a(\phi)$, $a\gwia(\phi)$, 
$\gamma_\mu$, $\lambda_\mu(T\oplus\zero)$ for all $\phi\in\Ka$, $|\mu|<1$ 
and bounded operators $T:\Ka\rightarrow\Ka$. We shall denote by 
$\A_{\Ka}(\Ha)$ the completion of $\A_{{\rm fin}\Ka}(\Ha)$ in the operator 
norm. 

\subsection{Normal ordering}
In algebras generated by (bosonic, fermionic or $q$-deformed) creation 
and annihilation operators one introduces normal or Wick's ordering where
in each expression one writes creation operators on the right hand side and 
 annihilation operators on the left hand side. Now we introduce an analogue 
of such ordering in algebras $\A\fin$. 
\begin{theo} Every element $S$ of an algebra $\A\fin$ can 
be written as a finite sum of products of the following form: an the left 
hand side creation operators, some $\lambda_\nu(T)$ operators, a 
$\gamma_\mu$ operator (for some $\mu$, $|\mu|\leq 1$) and on the right hand 
side annihilation operators. 
\begin{equation} S=\sum_i a\gwia(\phi_{i1})\cdots a\gwia(\phi_{in_i}) 
\lambda_{\nu_{i1}}(T_{i1}) \cdots \lambda_{\nu_{ik_i}} (T_{ik_i}) 
\gamma_{\mu_i} a(\psi_{i1})\cdots a(\psi_{im_i}).
\label{eq:kanoniczna} 
\end{equation}
\end{theo}
\begin{proof}
Note that an expression is in the mentioned form if and only if it does not 
contain any subexpression being the left hand side of one of equations 
(\ref{eq:komutacja1})---(\ref{eq:komutacja7}). If it does not hold by 
replacing the left hand side of appropriate equation by the right hand side 
we obtain an expression (or a sum of expressions) which is either shorter or 
has the same length but smaller number of disorderings. We can easily see 
that this procedure has to stop after finite number of iterations.
 \end{proof} 

By $\A^{(k,l)}$ we shall denote the completion of the space of 
operators that can be written in the normal ordering (\ref{eq:kanoniczna}) 
with exactly $k$ creation operators $a\gwia(\cdot)$ and $l$ annihilation 
operators $a(\cdot)$. Let 
$\A^{(k,\cdot)}=\overline{\bigoplus_{l\in{\N}}\A^{(k,l)}}$ and 
$\A^{(\cdot,l)}=\overline{\bigoplus_{k\in{\N}} \A^{(k,l)}} $.

For an integer number $n\in\Z$ we define $\A^{[n]}$ to be a completion of 
the space of operators that (not necessarily in the normal ordering) contain
exactly $n$ more creators than annihilators.
$\A^{[n]}=\overline{\bigoplus_{k,l\in \N,\ k-l=n} A^{(k,l)}}$.

For $S\in\A$, $n\in\Z$ let $S^{[n]}\in\A^{[n]}$ 
be a part of $S$ which contains exactly $n$ more creation than annihilation 
operators. More precisely, $S^{[n]}=\sum_{i\in\Z:i\geq 0, i+n\geq 0} 
\Pi_{i+n} S \Pi_i$. Note that $\|S^{[n]}\|\leq\|S\|$ because 
$$\|S^{[n]}\Psi\|^2= \sum_i \|\Pi_{i+n} S \Pi_i \Psi\|^2\leq\sum_i \|S \Pi_i 
\Psi\|^2\leq\|S\|^2 \sum_i \|\Pi_i \Psi\|^2=\|S\|\ \|\Psi\|^2.$$

\subsection{Extension of operators}
\begin{lem}\label{lem:extension} If $\Ha_1=\Ka\oplus{\cal L}_1$ and 
$\Ha_2=\Ka\oplus{\cal L_2}$ where ${\cal L_1}$, ${\cal L_2}$ are 
separable, infinite dimensional Hilbert spaces then there exists exactly one 
continuous $\ast$-isomorphism $V$ of Banach algebras $\A_{\Ka}(\Ha_1)$ and 
$\A_{\Ka}(\Ha_2)$ which maps operators $a(\phi), a\gwia(\phi), 
\lambda_\mu(T), \gamma_\mu\in \A_{\Ka}(\Ha_1)$ respectively on $a(\phi), 
a\gwia(\phi), \lambda_\mu(T), \gamma_\mu\in \A_{\Ka}(\Ha_2)$. Moreover, this 
$\ast$-isomorphism is an isometry. 

Particularly, if $\Ha_1\subset\Ha_2$ then this $\ast$-isomorphism assigns to 
an operator $S\in\A_{\Ka}(\Ha_1)$ its extension
$\tilde{S}\in\A_{\Ka}(\Ha_2)$ (since it does 
not lead to confusions we shall often denote both operators by the same 
letter). \end{lem}
\begin{proof} 
Let $U:\Ha_1\rightarrow\Ha_2$ be an
isometry such that $U$ limited to $\Ka$ is equal to identity. Such isometry 
exists because ${\cal L}_1$ and ${\cal L}_2$ have the same dimension.

We define now an second quantization of $U$, i.e.\ an isometry  
$\Gamma(U):\Gamma_{\Ha_1}\rightarrow\Gamma_{\Ha_2}$ by the 
formula $\Gamma(U)(\psi_1\otimes\cdots\otimes \psi_n)=U(\psi_1) \otimes 
\cdots \otimes U(\psi_n)$. 
The wanted $\ast$-isomorphism is $$\A_{\Ka}(\Ha_1)\ni S\mapsto \Gamma(U) 
 S \Gamma(U)\gwia\in\A_{\Ka}(\Ha_2).$$

The uniqueness of such isomorphism follows from the fact that it is uniquely 
defined on a dense subspace $\A\fin$ \end{proof} 
The lemma remains true if in the formulation we skip the assumption of 
separability, the proof of this fact is however more complicated.


\section{Stochastical integrals}
\label{sec:calki}
Since we are interested in stochastic calculus, since now we have 
$\Ha={\El}^2(\Rplus)$. We also introduce notation 
$\A=\A_\Ha$, $\Ha_t={\El}^2(0,t)$ and $\A_t=\A_{\Ha_t}$.

We shall investigate stochastic integrals with respect to four basic 
stochastic processes with values in algebra $\A$: annihilation 
$A(t)=a(\chi_{(0,t)})$, creation $A\gwia(t)=a\gwia(\chi_{(0,t)})$, 
gauge $\Lambda_\mu(t)=\lambda_\mu(\Pi_{(0,t)})$ and time $T(t)=t\jed$, where 
$\chi_I\in\Ha$ denotes a characteristic function of a set $I\subset\Rplus$ 
and
 $\Pi_I:\cal{L}^2(\Rplus)\rightarrow\cal{L}^2(I)$ denotes 
the orthogonal projection. 

\subsection{Bioperators and biprocesses}
If $S:\Rplus\rightarrow \A$ is a measurable function we shall call it 
a process. If for almost all $t\in \Rplus$ we have $S(t)\in \A_t$ we say it 
is adapted.

Elements of $\A\otimes\A$ will be called bioperators. A bioperator 
can be multiplied by an operator from left or right and the result is a 
bioperator: for $F,G,S\in \A$ we define $(F\otimes G) S=F \otimes (GS)$, 
$S(F\otimes G)=(SF)\otimes G$. Furthermore we define a ``musical'' product 
of a bioperator by an operator such that the result is an operator: 
$(F\otimes G)\sharp S=FSG$. We shall introduce a convolution: $(F\otimes 
G)\gwia=G\gwia\otimes F\gwia$.

If $R:\Rplus\rightarrow \A\otimes\A$ is a measurable 
function we shall call it a biprocess. If for almost all $t\in \Rplus$ 
we have $R(t)\in \A_t \otimes \A_t$ we say it is adapted, if for almost all 
$t\in \Rplus$
we have $R(t)\in \A_t \otimes \A$ or $R(t)\in \A \otimes \A_t$ we say that 
$R$ is respectively left- or right-adapted. 

A simple biprocess is a biprocess of 
the form $R(t)=\sum_{i=1}^{n} B_i \chi_{I_i}(t)$ where $B_i\in \A\otimes 
\A$ and $I_i$ are intervals.

\subsection{Stochastic integral of a simple biprocess}
A stochastic integral of a simple biprocess is defined as a Riemann sum
$$\int \left( \sum_{i=1}^{n} B_i \chi_{I_i}(t) \right)\sharp dS=
\sum_{i=1}^{n} B_i\sharp [S(s_i)-S(t_i)],$$ where $I_i=(s_i,t_i)$.

\subsection{Stochastic integrals with respect to the creation and 
annihilation process}
\label{subsec:nierownosci}
\subsubsection{Tensor product $\otimes_k$}
For $\phi, \psi_1,\dots, \psi_n\in\Ha$ we define a tensor product
$\otimes_k$ as
 $$\phi\otimes_k(\psi_1\otimes\cdots\otimes 
\psi_n)=
\left\{\begin{array}{c@{\quad:\quad}l} 
  \psi_1\otimes\cdots\otimes \psi_k\otimes \phi \otimes 
\psi_{k+1}\otimes\cdots \otimes\psi_n & k\leq n\\ 0 & k>n \end{array} 
\right. $$ and an operator $\jed\otimes_k 
P^{(n)}:\Hado{(n+1)}\rightarrow\Hado{(n+1)}$ which for $n\geq k$ is defined 
as 
$$[\jed\otimes_k P^{(n)}] 
(\psi_1\otimes\cdots\otimes\psi_k\otimes\phi\otimes 
\psi_{k+1}\otimes\cdots\otimes\psi_n)=$$ $$=\sum_{\sigma\in S_n} q^{{\rm 
inv}(\sigma)} 
\psi_{\sigma(1)}\otimes\cdots\otimes\psi_{\sigma(k)}\otimes\phi\otimes 
\psi_{\sigma(k+1)}\otimes\cdots\otimes\psi_{\sigma(n)} $$ 
and is a modification of $q$-deformed symmetrization operator, which does 
not move the factor on the $k+1$ position. For $n<k$ we 
take $\jed\otimes_k P^{(n)}=0$. 
\begin{lem} There exists a positive constant $\omega(q)$ such that for each 
$n$ we have
$$ P^{(n+1)} \leq \frac{1}{1-|q|}\ \jed \otimes P^{(n)},$$
$$ \jed \otimes P^{(n)} \leq \frac{1}{\omega(q)}\ P^{(n+1)}.$$
There exist positive constants $c_{k,q}$, $d_{k,q}$  
such that for each $n\geq k$ we have
$$P^{(n+1)} \leq c_{k,q}\ \jed \otimes_k P^{(n)},$$
$$\jed \otimes_k P^{(n)} \leq d_{k,q} P^{(n+1}.$$
\end{lem}
\begin{proof} Proof of the first two inequalities can be found in 
\cite{Boz}. Now we show the
third one.
$$P^{(n+1)}\leq\frac{1}{1-|q|}\jed\otimes 
P^{(n)}\leq\cdots\leq
\frac{1}{(1-|q|)^{k+1}}\jed^{\otimes(k+1)}\otimes 
P^{(n-k)}=$$ 
$$=\frac{1}{(1-|q|)^{k+1}}\jed\otimes_k [\jed^{\otimes 
k}\otimes P^{(n-k)}]\leq$$ 
$$\leq\frac{\omega(q)}{(1-|q|)^{k+1}}\jed\otimes_k [\jed^{\otimes 
(k-1)}\otimes P^{(n-k+1)}]\leq\cdots
\leq\frac{\omega(q)^{k-1}}{(1-|q|)^{k+1}}\jed\otimes_k P^{(n)}$$ 
The last inequality can be proven similarly. \end{proof}
\begin{coro} For any $\Psi\in\Gamma$ holds
$$\|\Psi\|_{\jed \otimes_k P}\leq \sqrt{d_{k,q}}\ \|\Psi\|.$$ 
If furthermore $\Psi\in\bigoplus_{n>k} \Hado{n}$ then
$$\|\Psi\|\leq \sqrt{c_{k,q}}\ 
\|\Psi\|_{\jed \otimes_k P}.$$ 
For every $\phi\in\Ha$ we have
$$\|a(\phi)\|=\|a\gwia(\phi)\|\leq\frac{\|\phi\|}{\sqrt{1-|q|}}.$$
\end{coro}
\begin{proof} The last inequality holds since
for each $\Psi\in\Gamma$ we have
$$\|a\gwia(\phi)\Psi\|\leq \frac{1}{\sqrt{1-|q|}} 
\|a\gwia(\phi)\Psi\|_{\jed\otimes P}=\frac{1}{\sqrt{1-|q|}}\|\phi\|\ 
\|\Psi\|.$$ \end{proof}

\subsubsection{Properties of $Q_k$}
Let $S\in\A_\Ka(\Ha)$ and let $\phi$ be a unital vector perpendicular to 
$\Ha$. By lemma \ref{lem:extension} there exists an operator 
$S:\Ha\oplus\phi\rightarrow\Ha\oplus\phi$ which is an extension of 
$S:\Ha\rightarrow\Ha$.

It is easy to see that for each $\Phi\in\Gamma_\Ha$ there exists an element 
of $\Gamma_\Ha$ denoted by $Q_k(S)\Phi$ such that
\begin{equation}\phi\otimes_k[Q_k(S) \Phi]=(\Pi_\phi\otimes_k \jed) S 
a\gwia(\phi)\Phi, \end{equation} 
where $\Pi_\phi$ denotes the orthogonal projection on the subspace spanned 
by $\phi$ and $\Pi_\phi\otimes_k \jed$ is an operator, which on tensor 
products of not more than $k$ vectors acts as $\zero$ and on longer tensor 
powers acts on the $(k+1)$-th factor by $\Pi_\phi$. 

Of course $Q_k(S):\Gamma\rightarrow\Gamma$ is a linear operator. We shall 
prove that $Q_k(S)$ is an element of the algebra $\A_\Ka(\Ha)$ and that this 
operator in the normal ordering has exactly $k$ creation operators, i.e.\
$Q_k(S)\in\A_\Ka^{(k,\cdot)}(\Ha)$.

Indeed, if $S$ is in the following form 
$$S=a\gwia(\phi_{1})\cdots a\gwia(\phi_{i})
\lambda_{\nu_{1}}(T_{1}) \cdots \lambda_{\nu_{l}} (T_{l}) \gamma_{\mu} 
a(\psi_{1})\cdots a(\psi_{j})$$
then a simple computation shows that
$$Q_k(S)=\left\{\begin{array}{c@{\quad:\quad}l} q^j \nu_1 \cdots \nu_l \mu S 
& k=i\\ 0 & k\neq i \end{array} \right. ,$$
and therefore $Q_k(S)\in \A^{(k,\cdot)}_\Ka{}\fin(\Ha)$. 

The general statement follows from the fact that 
$Q_k:\A_\Ka(\Ha)\rightarrow\A_\Ka(\Ha)$ is a continuous map:
\begin{equation} \label{eq:qkciagle} \left\|Q_k(S)\Psi\right\|=
\left\|\phi\otimes_k[Q_k(S)\Psi]\right\|_{\jed\otimes_k P}
=\left\|[\Pi_\phi\otimes_k \jed] S a\gwia(\phi) 
\Psi\right\|_{\jed\otimes_k P}
\leq \end{equation}
$$\leq \left\|S a\gwia(\phi) \Psi\right\|_{\jed\otimes_k P}
\leq \sqrt{d_{k,q}}\ \left\|S a\gwia(\phi) \Psi\right\|
\leq\sqrt{\frac{d_{k,q}}{1-|q|}}\ \left\|S\right\|\ \left\|\Psi\right\|.$$

In the following we shall often use the notation 
$Q_k:\A_\Ka(\Ha)\otimes\A_\Ka(\Ha)\rightarrow\A_\Ka(\Ha)$ defined on simple 
tensors by $Q_k(P\otimes R)=Q_k(P)R$.

\subsubsection{Norm of an integral with respect to the creation and 
annihilation processes}
\begin{theo} \label{theo:normacalkiagwia}
If $R:\Rplus\rightarrow \A\otimes \A$ is a simple left-adapted 
biprocess then
$$\left\|\int R(t)\sharp dA\gwia(t)\Psi\right\|\leq \sum_k\left(c_{k,q} \int 
\|(Q_k[R(t)]\Psi\|^2 dt\right)^{\frac{1}{2}}.$$
\end{theo}
\begin{proof}
Let $R(t)=\sum_i B_i \chi_{I_i}(t)$ where $I_i$ are disjoint intervals.
Since $R$ is left--adapted for any $\Psi\in\Gamma_{\Ha}$ we have
$$\sum_i B_i\sharp a\gwia(\chi_{I_i})\Psi=\sum_k \sum_i 
\chi_{I_i}\otimes_k[Q_k (B_i)\Psi],$$
therefore
$$\Bigl\| \sum_i B_i\sharp a\gwia(\chi_{I_i})\Psi \Bigr\|\leq \sum_k\
\Bigl\| \sum_i 
\chi_{I_i}\otimes_k [Q_k(B_i)\Psi]\Bigr\|,$$
$$\Bigl\|\sum_i \chi_{I_i}\otimes_k [Q_k(B_i)\Psi] \Bigr\|^2\leq
 c_{k,q}\ \Bigl\|\sum_i \chi_{I_i}\otimes_k [ Q_k(B_i)\Psi] 
\Bigr\|_{\jed\otimes_k P}^2=$$
$$=c_{k,q} \sum_i \left\|Q_k(B_i)\Psi\right\|^2\  
\left\|\chi_{I_i}\right\|^2=c_{k,q} \int\left\|Q_k[R(t)]\Psi\right\|^2\ dt$$
what proves claimed theorem. \end{proof}
Now we shall define appropriate seminorms on the space of biprocesses:
$$\|R\|_{A\gwia}=\sum_k \left(c_{k,q} \int_0^{\infty} 
\Bigl\|Q_k[R(t)]\Bigr\|^2 dt \right)^{\frac{1}{2}},$$
$$\|R\|_A=\|R\gwia\|_{A\gwia}.$$
\begin{theo} \label{theo:gestosc}
Simple adapted (respectively left-adapted or right-adapted) biprocesses are 
dense in the space of adapted (respectively left-adapted or right-adapted) 
biprocesses in seminorms $\|\cdot\|_A$ and $\|\cdot\|_{A\gwia}$.
\end{theo}
\begin{proof} Proof of an analogue fact can be found in the paper of 
Biane and Speicher \cite{BiS}. 
\end{proof}
Therefore we can define an integral with respect to the creation (or 
annihilation) process of a left--adapted (or resp.\ right--adapted) 
biprocess $R(t)$ with finite seminorm $\|\cdot\|_{A\gwia}$ (or 
$\|\cdot\|_A$) as a limit of integrals of a sequence of simple
biprocesses. 
 
We have the following \begin{theo} \label{theo:nierownoscagwia} If 
$R:\Rplus\rightarrow \A\otimes \A$ is a left-adapted biprocess and 
$\Psi\in\Gamma$ then $$\left\|\int R(t)\sharp 
dA\gwia(t)\right\|\leq\|R\|_{A\gwia},$$ $$\left\|\int R(t)\sharp 
dA\gwia(t)\Psi\right\|\leq \sum_k\left(c_{k,q} \int \|(Q_k[R(t)]\Psi\|^2 
dt\right)^{\frac{1}{2}}.$$ If $R:\Rplus\rightarrow \A\otimes \A$ is a 
right-adapted biprocess then $$\left\|\int R(t)\sharp 
dA(t)\right\|\leq\|R\|_{A}.$$ \end{theo}
%
%
%
%
\subsection{Stochastic integral with respect to the gauge process}
For an operator $S:\Gamma\rightarrow\Gamma$ and ${\cal V}$, ${\cal W}$ 
subspaces of $\Gamma$, we shall denote by $\|S\|_{{\cal V}\rightarrow{\cal 
W}}$ the operator norm of $S$ defined as
$$\|S\|_{{\cal V}\rightarrow{\cal W}}=\sup_{\Phi\in{\cal W},\ \|\Phi\|=1} \
\sup_{\Psi\in{\cal V},\ \|\Psi\|=1} | \langle \Phi,S \Psi \rangle |.$$

For $|\mu|<1$ introduce a gauge seminorm of a bioperator 
$B\in\A_t\otimes\A_t$ $$\|B\|_{\lambda_\mu}=\sup_{n,m\in\N} \sqrt{nm}\ 
\|B\sharp\lambda_\mu(\Pi_{(t,\infty)}) 
\|_{\Hado{n}\rightarrow\Hado{m}}$$ and if a biprocess $R$ is adapted we 
introduce its gauge seminorm as
$$\|R\|_{\Lambda_\mu}=\sup_{t \in \Rplus} \sup_{n,m\in\N} \sqrt{nm}\ 
\|R(t)\sharp\lambda_\mu(\Pi_{(t,\infty) })
\|_{\Hado{n}\rightarrow\Hado{m}}. $$
\begin{theo} \label{theo:estimation1}
If $R(t)$ is a simple adapted biprocess
then the following estimation holds:
$$\left\|\int R(t)\sharp d\Lambda_{\mu}(t)\right\|\op\leq 
\|R\|_{\Lambda_\mu}.$$ 
\end{theo}
As we shall see soon the assumption that $R$ is simple can be omitted.
\begin{proof}
Let us consider a Hilbert space $\Ha\oplus\widetilde{\Ha}$ such that 
there exists an operator
$U:\Ha\oplus\widetilde{\Ha}\rightarrow\Ha\oplus\widetilde{\Ha}$ which
restricted to $\Ha$ is a unitary operator $U:\Ha\rightarrow\widetilde{\Ha}$ 
and restricted to $\widetilde{\Ha}$ is equal to $\zero$. We 
introduce a process 
$\widetilde{\Lambda_\mu}(t)=\lambda_\mu(\Pi_{\widetilde{(0,t)}})$, 
where $\Pi_{\widetilde{I}}$ denotes the  orthogonal projection 
on the subspace $U[\cal{L}^2(I)]$ for a set $I\subset \Rplus$.

For any operator 
$X:\Ha\oplus\widetilde{\Ha}\rightarrow\Ha\oplus\widetilde{\Ha}$ we define 
$\lambda(X)=\lambda_1(X)$. Of course this operator is not bounded and 
therefore any manipulations with it have to be done carefully. Lemma 
\ref{lem:rownoscnorm} ensures that $\lambda(U)$ on $\Hado{n}$ is bounded 
and its norm equals to $\sqrt{n}$.

Let $\Gamma_\Ha\ni\Psi=\sum \Psi_n$, $\Gamma_\Ha\ni\Phi=\sum \Phi_n$, where
$\Psi_n,\Phi_n\in\Hado{n}$. For any measurable set $M$ we have
$$\left|\left\langle \Phi_n,\int_M R(t)\sharp 
d\Lambda_\mu(t) \Psi_m\right\rangle\right|=$$ $$=
\left|\left\langle\lamjed(U)\Phi_n,
\left( \int R(t)\sharp d\widetilde{\Lambda_{\mu}}(t) \right)
\lamjed(U\Pi_M)\Psi_m\right\rangle\right|\leq$$ 
$$\leq \sqrt{n}\ \|\Phi_n\|\  \|\lamjed(U\Pi_M)\Psi_m\|\ 
\left\|\int R(t)\sharp d\widetilde{\Lambda_{\mu}}(t)
\right\|_{\lamjed(U)[\Hado{m}]\rightarrow \lamjed(U)[\Hado{n}]}.$$
Let $R(t)=\sum_i B_i \chi_{I_i}(t)$ where $I_i$ are disjoint intervals.
For different values of $i$ operators 
$B_i\sharp\lambda_{\mu}(\Pi_{\widetilde{I_i}}): 
\lamjed(U)\Hado{m}\rightarrow \Gamma(\Ha\oplus\widetilde{\Ha})$ have 
mutually orthogonal images and cokernels. $$\Bigl\|\sum_i 
B_i\sharp\lambda_{\mu}(\Pi_{\widetilde{I_i}}) 
\Bigr\|_{\lamjed(U)\Hado{m}\rightarrow\lamjed(U)\Hado{n}}=$$ $$=\max_i\ 
\Bigl\|B_i\sharp\lambda_{\mu}(\Pi_{\widetilde{(t_i,\infty)}})
\Bigr\|_{\lamjed(U)\Hado{m}\rightarrow \lamjed(U)\Hado{n}}\leq$$
$$\leq\max_i\
\Bigl\|B_i\sharp\lambda_{\mu}(\Pi_{\widetilde{(t_i,\infty)}}) 
\Bigr\|_{(\Ha\oplus\widetilde{\Ha})^{\otimes 
m}\rightarrow(\Ha\oplus\widetilde{\Ha})^{\otimes n}}=$$ 
$$=\max_i\ 
\Bigl\|B_i\sharp\lambda_{\mu}(\Pi_{{(t_i,\infty)}}) 
\Bigr\|_{(\Ha\oplus\widetilde{\Ha})^{\otimes 
m}\rightarrow(\Ha\oplus\widetilde{\Ha})^{\otimes n}}=$$ 
$$=\max_i\ 
\Bigl\|B_i\sharp\lambda_{\mu}(\Pi_{(t_i,\infty)}) 
\Bigr\|_{\Ha^{\otimes m}\rightarrow\Ha^{\otimes n}},$$
where in the last equality we used the lemma \ref{lem:rownoscnorm} and in 
the last but one equality we used that second quantization 
$\Gamma(U+U\gwia):\Gamma(\Ha\oplus\widetilde{\Ha})\rightarrow 
\Gamma(\Ha\oplus\widetilde{\Ha})$ of unitary operator $U+U\gwia$ defined as 
$$\Gamma({U+U\gwia})(\phi_1 
\otimes\cdots\otimes\phi_n)=(U+U\gwia)(\phi_1) 
\otimes\cdots\otimes(U+U\gwia)(\phi_n)$$ 
 is again unitary and 
$$B_i\sharp\lambda_{\mu}(\Pi_{{\cal{L}^2(t_i,\infty)}})= 
\Gamma({U+U\gwia})
[B_i\sharp\lambda_{\mu}(\Pi_{\widetilde{(t_i,\infty)}}) ] 
\Gamma(U+U\gwia).$$ 

Hence
\begin{equation}\label{eq:silnazbieznoscgauge} \left| \left\langle \Phi, 
\int_M R(t)\sharp d\Lambda_{\mu}(t)\Psi\right\rangle \right|\leq 
\end{equation} 
$$ \sum_{n,m} \|\Phi_n\|\ \|\lamjed(U\Pi_M)\Psi_m\|\ 
\sqrt{n}\ \sup_i \Bigl\|B_i\sharp\lambda_{\mu} 
(\Pi_{I_i})\Bigr\|_{\Hado{m}\rightarrow \Hado{n}}\leq $$
$$\leq
\sum_{n,m} \|\Psi_n\|\ \|\Phi_m\|\ \sqrt{nm}\ \sup_i 
\Bigl\|B_i\sharp\lambda_{\mu} 
(\Pi_{I_i})\Bigr\|_{\Hado{m}\rightarrow \Hado{n}}\leq
\|\Psi\|\ \|\Phi\|\ \|R\|_{\Lambda_\mu}.$$
\end{proof}

We would like to extend the definition of a stochastic integral with 
respect to the gauge process to all biprocesses with finite seminorm 
$\|\cdot\|_{\Lambda_\mu}$ by taking the limit. However, since this seminorm 
is of $\El^{\infty}$ type, the space of simple biprocess is not dense in 
this space. However, we may have the pointwise convergence.
\begin{theo} 
For each adapted biprocess $R(t)$, $\|R\|_{\Lambda_\mu}<\infty$ there
exists a sequence of simple adapted biprocesses $R_i$, 
$\|R_i\|_{\Lambda_\mu}\leq \|R\|_{\Lambda_\mu}$ such that
$R_i(t)\longrightarrow R(t)$ (convergence in the seminorm 
$\|\cdot\|_{\lambda_\mu}$) for almost all $t$. \end{theo}
Proof of this theorem follows the well--known proofs in the classical theory 
of stochastic integration and we shall omit it.

\begin{theo} \label{theo:zbieznoscgauge}
If $(R_i)$ is a sequence of simple adapted biprocesses such that 
$\sup_i \|R_i\|_{\Lambda_\mu}<\infty$
and sequence $R_i(t)$ converges 
to some $R(t)$ in the seminorm $\|\cdot\|_{\lambda_\mu}$ then the sequence 
$\int R_i\sharp d\Lambda_\mu(t)$ converges in the strong operator topology.
\end{theo} 
\begin{proof} 
It is enough to prove that for each $\epsilon>0$ 
and all vectors $\Psi\in\Gamma$ 
$$\lim_{N\to \infty} \sup_{i,j>N} 
\|[R_i-R_j]\sharp d\Lambda_{\mu}(t)\Psi\| \leq \epsilon.$$

Let $M_{ij}=\{t:\|R_i(t)-R_j(t)\|_{\lambda_\mu}>\epsilon/2 \}$. 
We have
\[\int[R_i(t)-R_j(t)]\sharp d\Lambda_{\mu}(t)= \]
\begin{equation} \label{dwaskladniki}=
\int_{M_{ij}}[R_i(t)-R_j(t)]\sharp d\Lambda_{\mu}(t)+ \int_{\Rplus\setminus 
M_{ij}} [R_i(t)-R_j(t)]\sharp d\Lambda_{\mu}(t)\end{equation}
and the operator norm of the second summand does not exceed $\epsilon/2$.

We shall use the notation introduced in proof of the theorem 
\ref{theo:estimation1}.
It follows from the fact that $\bigcap_N \bigcup_{i,j>N} M_{ij}$ has measure 
$0$ that $$\lim_{N\rightarrow\infty} \sup_{i,j>N} \| \lamjed(U 
\Pi_{M_{ij}})\Psi_m\|=0$$ for any fixed vector $\Psi\in\Gamma$.

If we rewrite the inequality (\ref{eq:silnazbieznoscgauge}) replacing $M$ by 
$M_{ij}$ and $R(t)$ by $R_i(t)-R_j(t)$ we see by majorized convergence 
theorem that the first summand in (\ref{dwaskladniki}) tends strongly to 
$\zero$. \end{proof}

The preceding theorems allow us to extend the definition of an integral 
with respect to the gauge process to all adapted processed with finite 
norm $\|\cdot\|_{\Lambda_\mu}$ and to remove from the formulation of the 
theorem \ref{theo:estimation1} the assumption that the integrand is 
simple. 

\subsection{Integrals with respect to time} For a biprocess
$R$ we introduce its seminorm $$\|R\|_T=\int_0^{\infty} \|R\sharp\jed\|\ 
dt.$$ 
Of course we have $$\left\|\int R\sharp dT(t)\right\|\leq \|R\|_T.$$
%
%
%
%

\section{Iterated integrals}
\label{sec:iterowane}
\begin{lem} \label{lem:normalambda} 
If $R:\Rplus\rightarrow\A\otimes\A$ is a biprocess such that there exists 
an integer number $j$ such that $R:\Rplus\rightarrow\bigoplus_{|i|\leq 
j}\A^{[i]}$ then for any process $S:\Rplus\rightarrow\A$ we have 
$$\|RS\|_{\Lambda_\mu},\|SR\|_{\Lambda_\mu}\leq \sqrt{j+1}\ (2j+1)\ 
\sup_{t\in\Rplus} \|S(t)\|\op\ \|R\|_{\Lambda_\mu}$$ \end{lem} \begin{proof} 
$$\|SR\|_{\Lambda_\mu}\leq
\sum_{|i|\leq j} \sup_{t\in\Rplus} \sup_{n,m} \sqrt{nm}\ 
\Bigl\|S(t)^{[i]} R(t)\sharp 
\lambda_\mu(\Pi_{\Ka^{\perp}})\Bigr\|_{\Hado{n}\rightarrow\Hado{m}}\leq$$
$$\leq \sum_{|i|\leq j} \sup_{t\in\Rplus} \sup_{n,m} \sqrt{\frac{m}{m-i}}\ 
\sqrt{n(m-i)}\ \Bigl\|S(t)^{[i]}\Bigr\|\ \Bigl\|R(t)\sharp 
\lambda_\mu(\Pi_{\Ka^{\perp}})\Bigr\|_{\Hado{n}\rightarrow\Hado{(m-i)}}\leq$$
$$\leq \sqrt{j+1}\ (2j+1)\ \|S\|\ \|R\|_{\Lambda_\mu}$$
because $m-i\geq 1$ and $\frac{m}{m-i}\leq j+1$.
\end{proof}

\begin{lem} \label{lem:normaagwiazdka}
For any biprocess $R(t)$, process $S(t)$ and $\Psi\in\Gamma$ we have
$$\|RS\|_{A\gwia}\leq \|R\|_{A\gwia}\ \sup_{t\in\Rplus} \|S(t)\|$$
$$\left\|\int R(t)S(t)\sharp dA\gwia(t)\Psi\right\|\leq \|R\|_{A\gwia}\ 
\sup_{t\in\Rplus} \|S(t)\Psi\|$$
\end{lem}
\begin{proof}
It is enough to notice that $Q_k[R(t) S(t)]=Q_k[R(t)] S(t)$ and recall the 
definition of the norm $\|\cdot\|_{A\gwia}$ and 
theorem \ref{theo:nierownoscagwia} \end{proof}

\begin{lem} \label{lem:normaa} There exists a constant $C_{n,q}$ such that
if $R(t)$ is a biprocess such that 
$R(t)\in\bigoplus_{i\leq n}\A^{(i,\cdot)}\otimes\A$ and $S(t)$ is a
process such that $S(t)\in\bigoplus_{i\leq n}\A^{(i,\cdot)}$,  then
$$\|SR\|_{A\gwia} \leq C_{n,q} \|R\|_{A\gwia}\ \sup_t \|S(t)\|. $$

If $R$ is a biprocess such that $R(t)\in\bigoplus_{i\leq n}\A^{(i,\cdot)}$ 
and a sequence of processes $S_i(t)$ converges strongly to $\zero$ and 
fulfills $S_i(t)\in\bigoplus_{i\leq n}\A^{(i,\cdot)}$ and $\sup_i \sup_t 
\|S_i(t)\|<\infty$ then integrals $\int S_i(t) R(t) \sharp dA\gwia(t)$ 
converge strongly to $\zero$ as well.
 \end{lem} 
\begin{proof}
For a unital vector $\psi$ orthogonal to $\Ha$ we have (see inequality 
(\ref{eq:qkciagle})) 
$$\|Q_k[S(t)R(t)] \Psi\|\leq \sqrt{d_{k,q}}\ \|S(t) R(t)\sharp 
a\gwia(\phi)\Psi\|$$ 
$$\|R(t)\sharp a\gwia(\phi)\Psi\|=\biggl\|\sum_{l\leq 
n} \phi\otimes_l Q_l[R(t)] \Psi \biggr\|\leq \sum_{l\leq n} \sqrt{c_{l,q}}\ 
\|Q_l[R(t)]\|\ \|\Psi\| $$
and therefore for some constants $C_1,C_2,C_3$ which depend only on $q$ 
and $n$ we have $$\|Q_k[S(t)R(t)]\|\leq C_1\ \|S(t)\|\ \sum_{l\leq n} 
\|Q_l[R(t)]\|$$ $$\|Q_k[S(t)R(t)]\|^2 \leq C_2\ \|S(t)\|^2 \sum_{l\leq n} 
\|Q_l[R(t)]\|^2$$
$$\|SR\|_{A\gwia}=\sum_{k\leq 2n} \left(c_{k,q} \int \|Q_k[S(t)R(t)]\|^2 
dt \right)^{\frac{1}{2}}\leq$$
$$\leq C_3\ \sup_{t\in\Rplus} \|S(t)\| \left( \sum_{l\leq 
n} \int \|Q_l[R(t)]\|^2 dt \right)^{\frac{1}{2}}.$$
The second part of the lemma follows from the majorized convergence theorem.
\end{proof}

\begin{theo} \label{th:iterowane}
If
\begin{enumerate}
\item $\tau_n:\Rplus\rightarrow\Rplus$ is a sequence of measurable 
functions, $0\leq\tau_n(t)\leq t$ and functions $\tau_n(t)$ tend to $t$ 
uniformly, 
\item \label{zalozenie2} $S_1,S_2$ are processes, 
$S_1\in\{A\gwia,A,\Lambda_\mu,T\}$, $S_2\in\{A\gwia,A,\Lambda_\nu,T\}$ 
\item $R_1, R_2:\Rplus \rightarrow\A$ 
are adapted biprocesses and their appropriate norms of are finite: 
$\|R_1\|_{S_1}, \|R_2\|_{S_2}<\infty$,
\item \label{zal1} if 
$S_1=\Lambda_\mu$ then there exists $j$ such that for each $t\in\Rplus$ we 
have $\int_0^t R_2\sharp dS_2 \in\bigoplus_{|i|\leq j} \A^{[i]}$ \item 
\label{zal2} if $S_1=A$ then there exists $j$ such that for each 
$t\in\Rplus$ we have $R_1(t)\in \bigoplus_{i\leq j} 
\A\otimes\A^{(\cdot,i)}$ and $\int_0^t R_2\sharp dS_2\in\bigoplus_{i\leq 
j} \A^{(\cdot,i)}$ \item \label{zal3} if $S_2=\Lambda_\nu$ then there 
exists $j$ such that for each $t\in\Rplus$ we have $\int_0^t R_1\sharp 
dS_1 \in\bigoplus_{|i|\leq j} \A^{[i]}$ \item \label{zal4} 
\label{zalozenieostatnie} if $S_2=A\gwia$ then there exists $j$ such that 
for each $t\in\Rplus$ we have $R_2(t)\in \bigoplus_{i\leq 
j}\A^{(i,\cdot)}\otimes\A$ and $\int_0^t R_1\sharp dS_1\in\bigoplus_{i\leq 
j} \A^{(i,\cdot)}$
 \end{enumerate} then
$$\int_0^{\infty} R_1(t) \left[\int_0^t R_2(s) \sharp dS_2(s)\right] \sharp 
dS_1(t)=$$ 
$$=\lim_{n\rightarrow\infty} \int_0^{\infty} R_1(t) \left[\int_0^{\tau_n(t)} 
R_2(s) \sharp dS_2(s)\right] \sharp dS_1(t),$$
$$\int_0^{\infty} \left[\int_0^s R_1(t) \sharp dS_1(t)\right] R_2(s) \sharp 
dS_2(s)=$$ 
$$=\lim_{n\rightarrow\infty} \int_0^{\infty} \left[\int_0^{\tau_n(s)} 
R_1(t) \sharp dS_1(t)\right] R_2(s) \sharp dS_2(s).$$
\end{theo}
\begin{proof}
For $S_2\neq\Lambda_\nu$
functions $\left\|\int_{\tau_n(t)}^t R_2 \sharp 
dS_2(s)\right\|$ uniformly tend to $0$. Therefore by preceding 
lemmas appropriate norms of biprocesses $R_1(t) \left[\int_{\tau_n(t)}^t 
R_2(s) \sharp dS_2(s)\right]$ tend to $0$ what proves that the limit in 
the first equation holds in the operator norm. 

For $S_2=\Lambda_\nu$ and $S_1\in\{T,A\gwia\}$ for each 
$\Psi\in\Gamma$ functions $\left\|\int_{\tau_n(t)}^t R_2(s) \sharp dS_2(s) 
\Psi \right\|$ by theorem \ref{theo:zbieznoscgauge} uniformly tend to $0$ 
and theorem \ref{theo:nierownoscagwia} shows that the limit in the first 
equation holds in the strong operator topology.

For $S_1=\Lambda_\mu$ and $S_2=A\gwia$ lemma \ref{lem:normaa} and theorem 
\ref{theo:zbieznoscgauge} assure that the limit in the second equation 
holds in the strong operator topology.

For $S_1=\Lambda_\mu$, $S_2=\Lambda_\nu$ we introduce a Hilbert space 
$\Ha\oplus\widetilde{\Ha}\oplus\widehat{\Ha}$ such that there exist 
operators $U$, $V$ which restricted to $\widetilde{\Ha}\oplus\widehat{\Ha}$ 
are equal to $0$ and which map isometrically $\Ha$ respectively onto 
$\widetilde{\Ha}$ and $\widehat{\Ha}$. In the following for $I\subset\Rplus$
$\Pi_{\widetilde{I}}$ and $\Pi_{\widehat{I}}$ will denote the orthogonal 
projection respectively onto $U[\El^2(I)]$ and $V[\El^2(I)]$, furthermore 
$\widetilde{\Lambda_\mu}(t)=\lambda_\mu(\Pi_{\widetilde{(0,t)}})$ and 
$\widehat{\Lambda_\mu}(t)=\lambda_\mu(\Pi_{\widehat{(0,t)}})$. 

We have \begin{equation} \label{eq:iterowanelambdy} \int_0^{\infty} 
\left[\int_{\tau_n(s)}^s R_1(t) \sharp d\Lambda_\mu(t)\right] R_2(s) \sharp 
d\Lambda_\nu(s)=\end{equation}
 $$=\lamjed(V\gwia) \lamjed(U\gwia) 
\left[\int_0^\infty R_1(t) \sharp d\widetilde{\Lambda_\mu}(t)\right] 
\int_0^{\infty}
\lamjed(U\Pi_{(\tau_n(s),s)}) R_2(s) \sharp d\widehat{\Lambda_\nu}(s) 
\lamjed(V). $$ 
Note that even though $\lamjed(U)$, $\lamjed(V)$ are unbounded operators, 
the 
 right--hand side of this equation is well defined on domain 
$\Gamma(\Ha)$ (see proof of theorem \ref{theo:estimation1}).

For each $n$ we consider a sequence $t_{n,k}=k/n$ and a bounded operator 
$\Xi_n:\lamjed(U) \lamjed(V) \Gamma(\Ha)\rightarrow \lamjed(U) 
\lamjed(V) \Gamma(\Ha)$ defined as
 $$\Xi_n=\sum_{i} 
\lamjed(\Pi_{\widetilde{(u_{n,i},t_{n,i+1})}} )
\lamjed(\Pi_{\widehat{(t_{n,i},t_{n,i+1})}})  $$
where $u_{n,i}=\inf_{x\in(t_{n,i},t_{n,i+1})}\tau_n(x)$. It is easy to see 
that the sequence $(\Xi_n)$ strongly tends to $\zero$ and since
$$\int_0^{\infty}
\lamjed(\Pi_{\widetilde{(\tau_n(s),s)}}U) R_2(s) \sharp 
d\widehat{\Lambda_\nu}(s)\ \lamjed(V)=$$ $$=
\Xi_n \int_0^{\infty} 
\lamjed(\Pi_{\widetilde{(\tau_n(s),s)}}U) R_2(s) \sharp 
d\widehat{\Lambda_\nu}(s)\ \lamjed(V)$$
then the expression (\ref{eq:iterowanelambdy}) strongly tends to $\zero$, 
what proves the first equation.

All the other cases we obtain by taking adjoint of considered cases.
 \end{proof}

\section{It\^o's formula}
\label{sec:ito}
\subsection{Properties of $P_0$}
We introduce a map $P_0:\A_\Ka(\Ha)\rightarrow\A_\Ka(\Ha)$ as follows.
Let $\phi$ be a unital vector perpendicular to $\Ha$. $P_0(R)$ is an 
operator defined by
$$P_0(R)\Psi=a(\phi)R a\gwia(\phi)\Psi,$$
for all $\Psi\in\Gamma_\Ha$.
It is easy to see that for
$$R=a\gwia(\phi_{1})\cdots a\gwia(\phi_{i})
\lambda_{\nu_{1}}(T_{1}) \cdots \lambda_{\nu_{l}} (T_{l}) \gamma_{\mu} 
a(\psi_{1})\cdots a(\psi_{j})],$$
we have
$$P_0(R)=q^{i+j} \nu_1 \cdots \nu_l \mu R,$$
and therefore $P_0(R)\in\A_{\Ka}(\Ha)$.

\subsection{It\^o formula}
\begin{theo} If assumptions \ref{zalozenie2}---\ref{zalozenieostatnie} of 
theorem \ref{th:iterowane} are fulfilled 
then
\begin{equation} \label{eq:ito} \left[\int R_1(s)\sharp dS_1(s)\right] 
\left[\int R_2(t)\sharp dS_2(t)\right]= 
\int  [R_1(u)\sharp dS_1(u)]  [R_2(u)\sharp dS_2(u)]+\end{equation}
$$+\int  R_1(s) \left[\int_0^s R_2(t)\sharp dS_2(t)\right] \sharp 
dS_1(s)+\int \left[\int_0^t R_1(s)\sharp dS_1(s)\right] R_2(t)\sharp 
dS_2(t),$$ where the first summand on the right hand side is defined as 
follows: \begin{equation} \label{eq:kw0} \int  [R_1(u)\sharp dA(u)]  
[R_2(u)\sharp dA\gwia(u)]=\int (\jed\otimes P_0\otimes\jed)(R_1(u)R_2(u)) 
du,\end{equation} \begin{equation} \label{eq:kw1} \int [R_1(u)\sharp 
d\Lambda_\mu(u)]  [R_2(u)\sharp dA\gwia(u)]= \int [R_1(u)\sharp 
\gamma_\mu] R_2(u)\sharp dA\gwia(u),\end{equation} 
\begin{equation} \label{eq:kw2} \int [R_1(u)\sharp dA(u)]  
[R_2(u)\sharp d\Lambda_\nu(u)]= \int \left[R_1(u)[R_2(u)\sharp \gamma_\nu] 
\right]\sharp dA(u),\end{equation} 
\begin{equation} \label{eq:kw3} \int [R_1(u)\sharp 
d\Lambda_\mu(u)]  [R_2(u)\sharp d\Lambda_\nu(u)]= 
\int [R_1(u)\sharp \jed] R_2(u)\sharp 
d\Lambda_{\mu\nu}(u),\end{equation} 
\begin{equation}\int 
 [R_1(u)\sharp dS_1(u)] \label{eq:kw4} [R_2(u)\sharp dS_2(u)]=0\quad\quad 
\mbox{for other values of }S_1,S_2.\end{equation} Informally we may write it 
as follows: $$\begin{array}{cc}
  & dS_2 \\
dS_1 &
\begin{array}{|c||c|c|c|c|}
\hline
             & dA & dA\gwia & d\Lambda_\nu           & dT \\ 
\hline\hline
dA	& 0 & dT  & dA \gamma_\nu 
 & 0  \\ \hline dA\gwia   &  0    & 0 &   0 & 0  
\\ \hline d\Lambda_\mu & 0   &\gamma_\mu 
dA\gwia & d\Lambda_{\mu\nu} & 0 \\ \hline dT
 &  0  &  0  &  0  & 0 
\\ \hline \end{array} \\ \end{array}
$$ 
\end{theo}
\begin{proof}
For $n=1,2$ let $R_n=\sum R_{ni} \chi_{I_i}$ be simple adapted biprocesses. 
We assume that intervals $(I_i)$ form a partition, i.e.\ that they are 
disjoint. Note that we can replace partition $(I_i)$ by a refined partition 
$(I_i^{(\epsilon)})$ so that $\max_{i} |I^{(\epsilon)}_i|<\epsilon$.
\begin{equation}\left[\int R_1(s)\sharp dS_1(s)\right] \left[\int 
R_2(t)\sharp dS_2(t)\right]=\end{equation}
\begin{equation}\label{eq:kwantowy}=\sum_i [R_{1i}\sharp S_1(I_i)] 
[R_{2i}\sharp S_2(I_i)]+\end{equation}
\begin{equation}+\sum_j \sum_{i<j} [R_{1i}\sharp S_1(I_i)] [R_{2j}\sharp 
S_2(I_j)]+\sum_i \sum_{j<i} [R_{1i}\sharp S_1(I_i)] 
[R_{2j}\sharp S_2(I_j)]\end{equation}

The second and the third summands tend by theorem \ref{th:iterowane} to the 
second and the third summands of the right hand side of (\ref{eq:ito}).
We shall find the weak limit of the first summand when the grid of the 
partition tends to $0$.

If $S_1=T$ or $S_2=T$ then it is easy to see that the term 
(\ref{eq:kwantowy}) tends strongly to $\zero$.

If $S_1=A\gwia$ theorem \ref{theo:nierownoscagwia} gives 
$$\Bigl\|\sum_i [R_{1i} \sharp a\gwia(I_i)] 
[R_{2i}\sharp S_2(I_i)]\Psi\Bigr\|\leq$$
$$\leq\sum_k \left( c_{k,q} \sum_i \|Q_k(R_{1i}) [R_{2i}\sharp 
S_2(I_i)]\Psi\|^2 \ |I_i|\right)^{\frac{1}{2}}\leq$$
$$=\|R_1\|_{A\gwia} \ \sup_i \|R_{2i}\sharp S_2(I_i)\Psi\|,$$
what tends to $0$ as the grid of the partition $(I_i)$ tends to $0$.

By taking the adjoint we see that if $S_2=A$ then the term 
(\ref{eq:kwantowy}) weakly tends to $0$ as the grid of the partition $(I_i)$ 
tends to $0$.

The cases we have already considered show that the equation (\ref{eq:kw4}) 
holds.

If $S_1=A$ and $S_2=A\gwia$ then we can split the 
normally ordered form of the expression $[R_{1i}\sharp a(I_i)] 
[R_{2i}\sharp a\gwia(I_i)]$ into two parts: the first which does not 
contain operators $a\gwia(I_i)$, $a(I_i)$ and is equal to 
$|I_i|(\jed\otimes P_0\otimes\jed)[R_{1i} R_{2i}]$ and the second, which 
contains these operators in this order. The sum over $i$ of the second 
part tends in operator norm to $0$ as the grid of the partition tend to 
$0$ because it is in form (\ref{eq:kwantowy}) for $S_1=A\gwia$ and 
$S_2=A$, what proves equation (\ref{eq:kw0}).

If $S_1=\Lambda_\mu$ and $S_2=A\gwia$ then we can split the 
normally ordered form of the expression $[R_{1i}\sharp \lambda_\mu(I_i)] 
[R_{2i}\sharp a\gwia(I_i)]$ into two parts: one equal to $[R_{1i}\sharp 
\gamma_\mu] [R_{2i} \sharp a\gwia(I_i)]$ and the second part which 
contains operators $a\gwia(I_i)$, $\lambda_\mu(I_i)$ in this order. The 
sum over $i$ of the second part is in the form (\ref{eq:kwantowy}) with 
$S_1=A\gwia$ and $S_2=\Lambda_\mu$ therefore tends strongly to $0$ as the 
grid of partition tends to $0$, what proves equation (\ref{eq:kw1}).

By taking the adjoint of (\ref{eq:kw1}) we obtain the equation 
(\ref{eq:kw2}), i.e.\ the case $S_1=A$, $S_2=\Lambda_\nu$.

If $S_1=\Lambda_\mu$, $S_2=\Lambda_\nu$ we introduce a Hilbert space 
$\Ha\oplus\widetilde{\Ha}\oplus\widehat{\Ha}$ such that there exist 
operators $U$, $V$ which restricted to $\widetilde{\Ha}\oplus\widehat{\Ha}$ 
are equal to $0$ and which map isometrically $\Ha$ respectively onto 
$\widetilde{\Ha}$ and $\widehat{\Ha}$.
$$\Big\langle\Psi,\sum_i [R_{1i}\sharp \lambda_\mu(I_i)] 
[R_{2i}\lambda_\nu(I_i)]\Phi\Big\rangle=$$
$$=\bigg\langle\Psi,\lamjed (U\gwia) \lamjed(V\gwia) \left[ \int 
R_1(t)\sharp d\widetilde{\Lambda_\mu}(t) \right] \left[\int 
R_2(s)d\widehat{\Lambda_\nu}(s)\right] \times$$
$$\times \Big[\sum_{k} 
\lamjed(U\Pi_{I_k})\lamjed(V\Pi_{I_k})\Big]\Phi\bigg\rangle+ 
\Big\langle\Psi,\sum_i [R_{1i}\sharp \gamma_\mu] 
[R_{2i}\lambda_\nu(\Pi_{I_i})]\Phi\Big\rangle$$ It is easy to see that as 
the grid of the partition $(I_i)$ tends to $0$ that the operators $\sum_{k} 
\lamjed(U\Pi_{I_k})\lamjed(V\Pi_{I_k})$ strongly tend to $0$, therefore 
the first summand strongly tends to $0$,
what proves (\ref{eq:kw3}). 

Now it is enough to notice that any biprocesses can be approximated by 
simple biprocesses.
\end{proof}
Particularly, we can obtain a It\^o formula for 
noncommutative Brownian motion: $B(t)=A(t)+A\gwia(t)$
and Poisson process with intensively $l$ and 
deformation parameter $\mu$: $P_{\mu,l}(t)=\sqrt{l} 
(\gamma_\mu A(t)+A\gwia(t) 
\gamma_\mu)+\frac{\Lambda(t)}{\mu}+lt \gamma_\mu$ 
$$\begin{array}{c c}  &  dS_2 \\ dS_1 & 
\begin{array}{|c||c|c|} \hline
	& dt	& dB(t)	\\
\hline \hline
dt	& 0	& 0	\\
\hline
dB(t)	& 0	& dt	\\
\hline
\end{array} \\
\end{array}
\quad\quad
\begin{array}{c c}  &  dS_2 \\ dS_1 & 
\begin{array}{|c||c|c|}
\hline
	& dt	& dP_{\nu,l}(t)	\\
\hline \hline
dt	& 0	& 0	\\
\hline
dP_{\mu,l}(t)	& 0	& dP_{\mu\nu,l}(t)	\\
\hline 
\end{array} \\
\end{array} $$
\section{Final remarks}
In this paper we have presented foundations of $q$-deformed stochastic 
calculus. The lack of space does not allow us to present its applications, 
among them the connection between $q$-deformed stochastic integral and 
noncommtative local martingales. Especially interesting is the possibility 
of interpolation of classical Brownian motion and Poisson process by their 
bounded $q$-deformed analogues for $q\rightarrow 1$ where new tools are 
useful. There are also many questions concerning deformed Poisson process.

\section{Acknowledgements}
The author would like to thank prof.\ Marek Bo\.zejko for many inspiring 
discussions. This work was partially supported by the Scientific Research 
Committee in Warsaw under grant number P03A05415.


\begin{thebibliography}{99}
\bibitem [BiS]{BiS} P.Biane, R.Speicher, Stochastic calculus with respect to 
free Brownian motion and analysis on Wigner space. Preprint 1997, {\it 
Prob.\ Th.\ Rel.\ Fields} (in printing)
\bibitem [Bo\.z]{Boz} M.Bo\.zejko, Completely positive maps on Coxeter 
groups and the ultracontractivity of the q-Ornstein-Uhlenbeck semigroup,  
{\it Banach Center Publications} {\bf 43}, 87-93 (1998)
\bibitem [BS1]{BoS} M.Bo\.zejko, R.Speicher, Completely positive maps on 
Coxeter, deformed commutation relations, and operator spaces, {\it Math.\ 
Ann.}\ {\bf 300}, 97-120 (1994)
\bibitem [BS2]{BoS2} M.Bo\.zejko, R.Speicher, An example of a generalized 
Brownian motion II, {\it Quantum Probability and Related Topics VII}, 
ed.~L.Accardi, Singapore, World Scientific 1992, pp. 219-236
\bibitem [BSW1]{BSW1} B.Barnett, R.F.Streater, I.F.Wilde, The It\^o-Clifford 
integral, {\it J.\ Funct.\ Anal.\ }{\bf 48},172-212 (1982)
 \bibitem [BSW2]{BSW2} B.Barnett, R.F.Streater, I.F.Wilde, Quasi-free 
quantum stochastic integrals for the CAR and CCR, {\it J.\ Funct.\ Anal.\ 
}{\bf 52}, 19-47 (1983)
\bibitem [FB]{FB} U.Frisch, R.Bourret, Parastochastics, {\it J.\ 
Math.\ Phys.\ }{\bf 11}, 364-390 (1970) 
\bibitem [HuP]{HuP} R.L.Hudson, 
K.R.Parthasarathy, Quantum It\^o's formula and stochastic evolution, {\it 
Comm.\ Math.\ Phys.\ }{\bf 93} 1994, 301-323 
\bibitem [K\"uS]{KuS} 
B.K\"ummerer, R.Speicher, Stochastic integration on the Cuntz algebra 
$O_\infty$, {\it J.\ Funct.\ Anal.\ }{\bf 103}, 372-408 (1992)
\bibitem [Spe]{Spe} R.Speicher, Stochastic Integration on the Full Fock 
Space with the Help of a Kernel Calculus, Publications of the Research 
Institute for Mathematical Sciences, Kyoto University, {\bf 27}, No. 1 (1991)
\end{thebibliography}
\end{document}